\documentclass{article}
\usepackage{amssymb}
\usepackage{graphicx}

\input{tcilatex}

\begin{document}

\title{Quantum Stackelberg duopoly in the presence of correlated noise}
\author{Salman Khan\thanks{%
sksafi@phys.qau.edu.pk}, M. Ramzan\thanks{%
mramzan@phys.qau.edu.pk}, M. Khalid Khan\thanks{%
mkkhan@qau.edu.pk} \\
Department of Physics, Quaid-i-Azam University, \\
Islamabad 45320, Pakistan.}
\maketitle

\begin{abstract}
We study the influence of entanglement and correlated noise using correlated
amplitude damping, depolarizing and phase damping channels on the quantum
Stackelberg duopoly. Our investigations show that under the action of
amplitude damping channel a critical point exists for unentangled initial
state as well, at which firms get equal payoffs. The game becomes a follower
advantage game when the channel is highly decohered. Two critical points
corresponding to two values of the entanglement angle are found in the
presence of correlated noise. Within the range of these limits of
entanglement angle, the game is follower advantage game. In case of
depolarizing channel, the payoffs of the two firms are strongly influenced
by the memory parameter. The presence of quantum memory ensures the
existence of Nash equilibrium for the entire range of decoherence and
entanglement parameters for both the channels. A local maximum in the
payoffs is observed which vanishes as the channel correlation increases.
Moreover, under the influence of depolarizing channel, the game is always a
leader advantage game. Furthermore, it is seen that phase damping channel
does not effect the outcome of the game.\newline
\end{abstract}

\section{Introduction}

Game theory is the mathematical study of interaction among independent, self
interested agents. It emerged from the work of Von Neumann \cite{von neumann}%
, and is now used in various disciplines like economics, biology, medical
sciences, social sciences and physics \cite{Piotrowski,Baaquie}. Due to
dramatic development in quantum information theory \cite{Neilson}, the game
theorists [$5$-$12$] have made strenuous efforts to extend the classical
game theory into the quantum domain. The first attempt in this direction was
made by Meyer \cite{Meyer} by quantizing a simple coin tossing game.
Applications of quantum games are reviewed by several authors [$14$-$18$]. A
formulation of quantum game theory based on the Schmidt decomposition is
presented by Ichikawa et al. \cite{Taksu2}. Recently, Xia et al. \cite%
{Xia1,Xia2} have investigated the quantum Stackelberg duopoly game under the
influence of decoherence and have found a critical point for the maximally
entangled initial state against the damping parameter for the amplitude
damping environment under certain conditions.

In practice no system can be fully isolated from its environment. The
interaction between system and environment leads to the destruction of
quantum coherence of the system. It produces an inevitable noise and results
in the loss of information encoded in the system \cite{Zurek}. This gives
rise to the phenomenon of decoherence. Quantum information are encoded in
qubits during its transmission from one party to another and require
communication channels. In a realistic situation, the qubits have a
nontrivial dynamics during transmission because of their interaction with
the environment. Therefore, a party may receive a set of distorted qubits
because of the disturbing action of the channel. Studies on quantum channels
have attracted a lot of attention in recent past \cite{Palma,Skeen}. Early
work in this direction was devoted mainly to memoryless channels for which
consecutive signal transmissions through the channel were not correlated. In
the correlated channels (i.e. the channels with memory), the noise acts on
consecutive uses of the channel. The effect of decoherence and correlated
noise in quantum games have produced interesting results and is studied by
number of authors \ [$23$-$28$].

In this paper, we study the effect of correlated noise introduced through
amplitude damping, phase damping and depolarizing channels parameterized by
the decoherence parameters $p_{1}$ and $p_{2}$ and the memory parameters $%
\mu _{1}$ and $\mu _{2}$, on the quantum Stackelberg duopoly game. The
decoherence parameters $p_{i}$ and the memory parameters $\mu _{i}$ range
from $0$ to $1$. The lower and upper limits of decoherence parameter $p_{i}$%
\ correspond to the undecohered and fully decohered cases, respectively.
Whereas the lower and upper limits of memory parameter $\mu _{i}$\
correspond to the uncorrelated and fully correlated cases, respectively. It
is seen that there exists a critical point in the case of amplitude damping
channel for initially unentangled state\textbf{\ }at which both firms have
equal payoffs. The game transforms from leader advantage to the follower
advantage game beyond this point, for highly decohered case in the presence
of memory. For initially entangled state under the influence of amplitude
damping channel we found two critical points. The game behaves as a follower
advantage game within these two critical points. In the case of depolarizing
channel the high correlation results in high payoffs. However, phase damping
channel has no effect on the game dynamics.

\section{Stackelberg duopoly game}

Stackelberg duopoly is a market game, which is rather different from the
Cournot duopoly game. In Cournot duopoly game, two firms simultaneously
provide a homogeneous product to the market and guess that what action the
opponent will take. However Stackelberg duopoly is a dynamic model of
duopoly game in which one firm, say firm $A$ moves first and the other firm,
say $B$, goes after. Before making its decision, firm $B$ observes the move
of firm $A$. This transforms the static nature of Cournot duopoly game to a
dynamic one. Firm $A$ is usually called the leader and firm $B$ the
follower, on this basis the game is also called the leader-follower model
\cite{Gibbons}. In classical Stackelberg duopoly it is assumed that firm $B$
will respond optimally to the strategic decision of firm $A$. As firm $A$
can precisely predict firm $B$'s strategic decision, firm $A$ chooses its
move in such a way that maximizes its own payoff. This informational
asymmetry makes the Stackelberg duopoly as the first mover advantage game.

A number of authors have proposed various quantization protocols for
observing the behavior of Stackelberg duopoly game in the quantum realm
[9,29-32]. It has been shown that quantum entanglement affects payoff of the
first mover and produces an equilibrium that corresponds to classical static
form of the same game \cite{Iqbal2}. The effects of decoherence produced by
various prototype quantum channels on quantum Stackelberg duopoly have been
studied by Zhu et al. \cite{Xia1}. We study the effects of correlated noise
on quantum Stackelberg duopoly, using amplitude damping, phase damping and
depolarizing channels.

\section{Calculations}

In a quantum Stackelberg duopoly game, for each firm $A$ and $B$ the game
space is a two dimensional complex Hilbert space of basis vectors $|0\rangle
$ and $|1\rangle $, that is, the game consists of two qubits, one for each
firm. We consider that the initial state of the game is given by%
\begin{equation}
|\psi _{i}\rangle =\cos \theta |00\rangle +\sin \theta |11\rangle  \label{E1}
\end{equation}%
where $\theta $ is a measure of entanglement. The state is maximally
entangled at $\theta =\frac{\pi }{4}$. In the presence of noise the
evolution of an arbitrary system can be described in terms of Kraus
operators as \cite{Neilson}%
\begin{equation}
\rho =\sum\limits_{l}E_{l}\rho _{i}E_{l}^{\dag }  \label{1}
\end{equation}%
where $\rho _{i}=|\psi _{i}\rangle \langle \psi _{i}|$ is the initial
density matrix and the Kraus operators $E_{l}$ satisfy the following
completeness relation%
\begin{equation}
\sum\limits_{l}E_{l}^{\dag }E_{l}=1  \label{2}
\end{equation}%
The single qubit Kraus operators for uncorrelated quantum amplitude damping
channel are given as \cite{Palma,Skeen}%
\begin{equation}
E_{0}=\left(
\begin{array}{cc}
1 & 0 \\
0 & \sqrt{1-p}%
\end{array}%
\right) ,\mathrm{\qquad }E_{1}=\left(
\begin{array}{cc}
0 & \sqrt{p} \\
0 & 0%
\end{array}%
\right)  \label{3}
\end{equation}%
The Kraus operators for amplitude damping channel with correlated noise for
a two qubit system are given as \cite{Skeen}%
\begin{equation}
E_{00}^{c}=\left(
\begin{array}{cccc}
\sqrt{1-p} & 0 & 0 & 0 \\
0 & 1 & 0 & 0 \\
0 & 0 & 1 & 0 \\
0 & 0 & 0 & 1%
\end{array}%
\right) ,\mathrm{\qquad }E_{11}^{c}=\left(
\begin{array}{cccc}
0 & 0 & 0 & 0 \\
0 & 0 & 0 & 0 \\
0 & 0 & 0 & 0 \\
\sqrt{p} & 0 & 0 & 0%
\end{array}%
\right)  \label{5}
\end{equation}%
The action of such a channel on the the initial density matrix \ of the
system is given by%
\begin{equation}
\rho =\left( 1-\mu \right) \sum\limits_{m,n=0}^{1}E_{mn}^{u}\,\rho
_{i}\,E_{mn}^{u\dag }+\mu \sum\limits_{l=0}^{1}E_{ll}^{c}\,\rho
_{i}\,E_{ll}^{c\dag }  \label{6}
\end{equation}%
where the superscripts $u$\ and $c$\ represent the uncorrelated and
correlated parts of the channel, respectively. The above relation means that
with probability $\mu $ the noise is correlated and with probability $\left(
1-\mu \right) $ it is uncorrelated. The Kraus operators for phase damping
channel with uncorrelated noise for a system of two qubits are given as \cite%
{Palma,Skeen}%
\begin{equation}
E_{mn}^{u}=\sqrt{e_{m}e_{n}}\sigma _{m}\otimes \sigma _{n},\hspace{0.5in}%
m,n=0,\mathrm{{\ }3}  \label{7}
\end{equation}%
whereas the one with correlated noise are given as%
\begin{equation}
E_{ll}^{c}=\sqrt{e_{l}}\sigma _{l}\otimes \sigma _{l},\mathrm{{\hspace{0.5in}%
}l=0,{\ }3}  \label{8}
\end{equation}%
\textbf{\ }Similarly, the Kraus operators for depolarizing channel are
described by equations (\ref{7}) and (\ref{8}) with indices run from $0$\ to
$3$. Where $\sigma _{0}$ is identity operator for a single qubit and $\sigma
_{1},\sigma _{2}$ and $\sigma _{3}$\ are the Pauli spin operators. For phase
damping channel, $e_{0}=\left( 1-p_{i}\right) $\ and $e_{3}=p_{i}$ and for
depolarizing channel $e_{0}=\left( 1-p_{i}\right) $,\ $e_{1}=e_{2}=e_{3}=%
\frac{1}{3}p_{i}$, where $p_{i}$ correspond to the decoherence parameters of
the first and second use of the channel. The action of such a channel on the
quantum system can be defined in a similar fashion as described earlier in
equation (\ref{6}).

In quantum Stackelberg duopoly game each firm has two possible strategies $I$%
, the identity operator and $C$, the inversion operator or Pauli's bit-flip
operator. Let $x$ and $1-x$ stand for the probabilities of $I$ and $C$ that
firm $A$ applies and $y$, $1-y$, are the probabilities that firm $B$
applies, respectively. The final state after the action of the channel is
given by \cite{Marrinatto2}%
\begin{eqnarray}
\rho _{f} &=&xyI_{A}\otimes I_{B}\ \rho \ I_{A}^{\dag }\otimes I_{B}^{\dag
}+x\left( 1-y\right) I_{A}\otimes C_{B}\ \rho \ I_{A}^{\dag }\otimes
C_{B}^{\dag }  \nonumber \\
&&+y\left( 1-x\right) C_{A}\otimes I_{B}\ \rho \ C_{A}^{\dag }\otimes
I_{B}^{\dag }  \nonumber \\
&&+\left( 1-x\right) \left( 1-y\right) C_{A}\otimes C_{B}\ \rho \
C_{A}^{\dag }\otimes C_{B}^{\dag }  \label{9}
\end{eqnarray}%
where $\rho $ (equation (\ref{6})) is the density matrix of the game after
the channel action.

Suppose that the player's moves in the quantum Stackelberg duopoly are given
by probabilities lying in the range $[0,1]$. In classical duopoly game the
moves of firms $A$ and $B$ are given by quantities $q_{1}$ and $q_{2}$,
which have values in the range $[0,\infty )$. We assume that firms $A$ and $%
B $ agree on a function that uniquely defines a real positive number in the
range $(0,1]$ for every quantity $q_{1}$, $q_{2}$ in $[0,\infty )$. Such a
function is given by $1/(1+q_{i})$, so that firms $A$ and $B$ find $x$ and $%
y $, respectively, as
\begin{equation}
x=\frac{1}{1+q_{1}}\mathrm{,\qquad }y=\frac{1}{1+q_{2}}  \label{10}
\end{equation}%
The payoffs of firms $A$ and $B$ are given by the following trace operations%
\begin{equation}
P_{A}\left( q_{1},q_{2}\right) =\mathrm{Tr}\left[ \rho _{f}P_{A}^{\mathrm{op}%
}\left( q_{1},q_{2}\right) \right] \mathrm{,\qquad }P_{B}\left(
q_{1},q_{2}\right) =\mathrm{Tr}\left[ \rho _{f}P_{B}^{\mathrm{op}}\left(
q_{1},q_{2}\right) \right]  \label{11}
\end{equation}%
where $P_{A}^{\mathrm{op}}$, $P_{B}^{\mathrm{op}}$ are payoff operators of
the firms and are given by%
\begin{eqnarray}
P_{A}^{\mathrm{op}}\left( q_{1},q_{2}\right) &=&\frac{q_{1}}{q_{12}}\left(
k\rho _{11}-\rho _{22}-\rho _{33}\right)  \nonumber \\
P_{B}^{\mathrm{op}}\left( q_{1},q_{2}\right) &=&\frac{q_{2}}{q_{12}}\left(
k\rho _{11}-\rho _{22}-\rho _{33}\right)  \label{12}
\end{eqnarray}%
where $\rho _{ii}$ are the diagonal elements of the final density matrix, $k$
is a constant as given in ref. \cite{Gibbons} and $q_{12}$ is given by%
\begin{equation}
q_{12}=\frac{1}{\left( 1+q_{1}\right) \left( 1+q_{2}\right) }  \label{13}
\end{equation}

The backward-induction outcome in the Stackelberg duopoly is found by first
finding the reaction $R_{2}\left( q_{1}\right) $ of firm $B$ to an arbitrary
quantity $q_{1}$ chosen by firm $A$. It is found by differentiating firm $B$%
's payoff with respect to $q_{2}$, and maximizing the result for $q_{1}$ and
can be written as%
\begin{equation}
R_{2}\left( q_{1}\right) =\max P_{B}\left( q_{1},q_{2}\right)  \label{14}
\end{equation}%
Once firm $B$ chooses this quantity, firm $A$ can compute its optimization
problem by differentiating its own payoff with respect to $q_{1}$and then
maximizing it to find the value $q_{1}=q_{1}^{\ast }$. Using the value of $%
q_{1}^{\ast }$ in equation (\ref{14}) we can get the value of $q_{2}^{\ast }$%
. These quantities define the backward-induction outcome of quantum
Stackelberg duopoly game and represent the subgame perfect Nash equilibrium
point. The payoffs of the firms at this point can be found using equation (%
\ref{11}).

\section{Results and discussion}

We suppose that before the firms measure their payoffs the game evolves
twice through different quantum correlated channels. That is, the state,
prior and after the firms apply their operators, is influenced by the
correlated noisy channels.

\subsection{Correlated amplitude damping channel}

The subgame perfect Nash equilibrium point for the game under the influence
of correlated quantum amplitude damping channel becomes%
\begin{eqnarray}
q_{1}^{\ast } &=&\frac{-k\cos ^{2}\theta +A_{1}\left( p_{1},p_{2},\mu
_{1},\mu _{2}\right) }{-4+A_{2}\left( p_{1},p_{2},\mu _{1},\mu _{2}\right) }
\nonumber \\
q_{2}^{\ast } &=&\frac{\frac{1}{4}k\cos ^{2}\theta -B_{1}\left(
p_{1},p_{2},\mu _{1},\mu _{2}\right) }{16+B_{2}\left( p_{1},p_{2},\mu
_{1},\mu _{2}\right) }  \label{15}
\end{eqnarray}%
where the damping function $A_{i}$ and $B_{i}$ are given in appendix A.

If we consider the influence of decoherence in the second evolution only ($%
p_{1}=0$), equation (\ref{15}) for unentangled initial state reduces to the
following form
\begin{eqnarray}
q_{1}^{\ast } &=&\frac{1}{2-4p_{2}(1-\mu _{2})}  \nonumber \\
q_{2}^{\ast } &=&\frac{1-2p_{2}(1-\mu _{2})}{2[2-p_{2}\{6-5\mu
_{2}-p_{2}(1-\mu _{2})(5-8\mu _{2})\}]}  \label{16}
\end{eqnarray}%
Here we have taken $k=1$. The firms' payoffs under this situation become%
\begin{eqnarray}
P_{A} &=&\frac{1}{8-16p_{2}(1-\mu _{2})}  \nonumber \\
P_{B} &=&\frac{1-2p_{2}(1-\mu _{2})}{8\{2-p_{2}(6-5\mu _{2})+p_{2}^{2}(8\mu
_{2}+5)(\mu _{2}-1)\}}  \label{17}
\end{eqnarray}%
In the classical form of the duopoly the perfect game Nash equilibrium is a
point, whereas in this case it is a function of both decoherence and memory
parameters. It can be easily seen that the results of ref. \cite{Xia1} are
retrieved by setting $\mu _{2}=0$, and setting $p_{2}=0$ reproduces the
results of classical game. The existence of Nash equilibrium requires that
firms' moves ($q_{1}^{\ast }$ and $q_{2}^{\ast }$) should have positive
values. It can easily be checked using equation (\ref{16}) that in the
absence of quantum memory, no Nash equilibrium exists for $p_{2}>\frac{1}{2}$%
. The presence of quantum memory allows the existence of Nash equilibrium
for the entire range of values of $p_{2}$, when $\mu _{2}\geq 0.85$.
\begin{figure}[h]
\begin{center}
\begin{tabular}{ccc}
\vspace{-0.5cm} \includegraphics[scale=0.4]{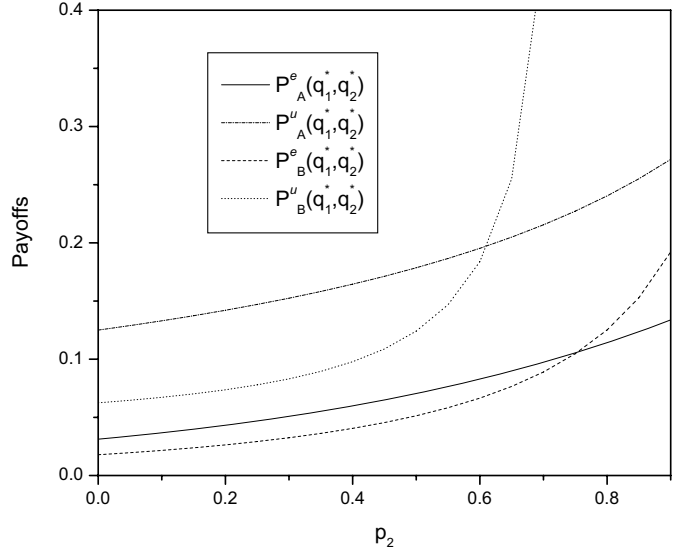}\put(-350,220) &  &
\end{tabular}%
\end{center}
\caption{The payoffs ($P_{A}(q_{1}^{\ast },q_{2}^{\ast })$ and $%
P_{B}(q_{1}^{\ast },q_{2}^{\ast })$) are plotted at the subgame perfect Nash
equilibrium point against the decoherence parameter $p_{2}$ both for
entangled and unentangled initial states under the action of amplitude
damping channel, with values of other parameters $k=1$ and $p_{1}=\protect%
\mu _{1}=0$. For maximally entangled initial state the memory parameter $%
\protect\mu _{2}=0.9$, and for unentangled initial state $\protect\mu %
_{2}=0.7$}
\end{figure}
To see the influence of decoherence and quantum memory on firms' payoffs at
the subgame perfect Nash equilibrium, we plot the payoffs (equation (\ref{17}%
)) in figure $1$ as a function of decoherence parameter $p_{2}$. In the
figure the dotted and dashed-dotted lines represent firm $A$ and firm $B$
payoffs for unentangled initial state, respectively. The superscripts $u$
and $e$ of $P_{A}$ ($P_{B}$) in the figure stand for unentangled and
entangled initial states respectively. It can be seen from the figure that a
critical point exists due to the presence of memory at which both firms are
equally benefited. This situation has not been observed, in the absence of
memory, for unentangled initial state of the game. That is, in the absence
of quantum memory the game is always first mover advantage game. For highly
decohered channel and $\mu <1$, a transition from first mover advantage into
second mover advantage occurs in the game behavior. It can also be seen from
equation (\ref{17}) that for fully correlated and fully decohered channel,
both firms are equally benefited and get a payoff equal to $\frac{1}{8}$. It
can also be shown that for smaller values of decoherence, the game is always
first mover advantage irrespective of the value of the quantum memory.

For a maximally entangled initial state of the game, the subgame perfect
Nash equilibrium point becomes%
\begin{eqnarray}
q_{1}^{\ast } &=&\frac{1-3p_{2}^{2}(-1+\mu _{2})-p_{2}(2-3\mu _{2})}{%
4-8p_{2}(1-\mu _{2})}  \nonumber \\
q_{2}^{\ast } &=&\frac{%
\begin{array}{c}
\lbrack \lbrack -1+p_{2}\{2+3p_{2}(-1+\mu _{2})-3\mu _{2}\}] \\
\times \{1+2p_{2}(-1+\mu _{2})\}]%
\end{array}%
}{%
\begin{array}{c}
\lbrack -7+p_{2}(28-26\mu _{2})+9p_{2}^{4}(-1+\mu _{2})^{2}+p_{2}^{2} \\
\times (-22+46\mu _{2}-23\mu _{2}^{2})-6p_{2}^{3}(2-5\mu _{2}+3\mu _{2}^{2})]%
\end{array}%
}  \label{19}
\end{eqnarray}%
The firms' payoffs corresponding to these values of $q_{i}^{\ast }$ become%
\begin{eqnarray}
P_{A} &=&\frac{\{-1+p_{2}(2-3\mu _{2})+3p_{2}^{2}(-1+\mu _{2})\}^{2}}{%
32(1+2p_{2}(-1+\mu _{2}))}  \nonumber \\
P_{B} &=&\frac{%
\begin{array}{c}
\lbrack -\{1+2p_{2}(-1+\mu _{2})\}\{-1+p_{2}(2-3\mu _{2}) \\
+3p_{2}^{2}(-1+\mu _{2})\}^{2}]%
\end{array}%
}{%
\begin{array}{c}
\lbrack 8\{-7+p_{2}(28-26\mu _{2})+9p_{2}^{4}(-1+\mu _{2})^{2} \\
+p_{2}^{2}(-22+46\mu _{2}-23\mu _{2}^{2})-6p_{2}^{3}(2-5\mu _{2}+3\mu
_{2}^{2})\}]%
\end{array}%
}  \label{20}
\end{eqnarray}%
One can easily check that these results reduce to the results of refs. \cite%
{Xia1, Iqbal2} by setting $\mu _{2}=0$ and $p_{2}=0$ respectively. The
payoffs of firms for the maximally entangled initial state are plotted as
function of decoherence parameter $p_{2}$ in figure $1$. The solid line
represents firm $A$ payoff and the dashed line represents firm $B$ payoff
for maximally entangled initial state. One can easily check that for a
maximally entangled initial state, the presence of quantum memory makes the
firms better off as compared to the uncorrelated case of the channel.
Furthermore, quantum memory shifts the critical point to higher payoffs than
the critical point for maximally entangled initial state when the channel is
uncorrelated. This can also be shown that for a given decoherence level, the
firms become worse off and the game becomes a follower advantage as the
channel becomes more correlated.
\begin{figure}[h]
\begin{center}
\begin{tabular}{ccc}
\vspace{-2cm} \includegraphics[scale=0.4]{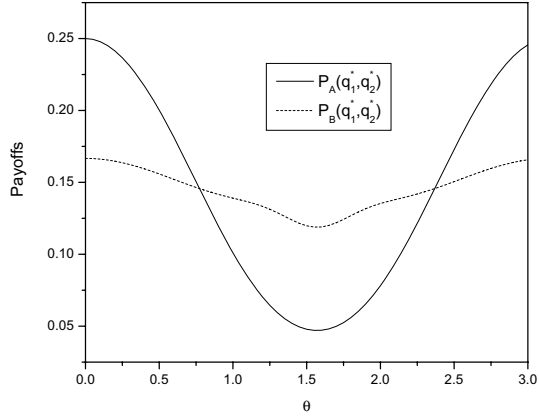}\put(-350,220) &
&
\end{tabular}%
\end{center}
\caption{The payoffs of the two firms are plotted at the subgame perfect
Nash equilibrium point against the entanglement angle $\protect\theta $
under the action of amplitude damping channel, with other parameters $%
p_{1}=p_{2}=0.5$, $k=1$ and $\protect\mu _{1}=\protect\mu _{2}=0.5$.}
\end{figure}
The effect of entanglement in the initial state on the firms' payoffs is
shown in figure $2$\textbf{. }Here we have taken into account the two
decohering correlated processes. It can be checked that for certain range of
decoherence parameters, in the absence of quantum memory, the moves of firms
are negative. This means that no Nash equilibrium exists in the region of
these values of decoherence parameters. The presence of quantum memory,
however, resolves this problem. As can be seen from figure $2$, the presence
of quantum memory in the payoff functions results into two critical points
corresponding to two different values of entanglement parameter. The game is
a follower advantage in this range of entanglement angle. The leader firm is
worse off in this range of values of $\theta $ and a global minimum in
payoffs occurs at $\theta =\frac{\pi }{2}$.

\subsection{Correlated Depolarizing channel}

When the game evolves under the influence of correlated depolarizing
channel, the moves of firms at subgame perfect Nash equilibrium point become
\begin{eqnarray}
q_{1}^{\ast } &=&\frac{81k\cos ^{2}\theta +A_{1}^{\prime }\left(
p_{1},p_{2},\mu _{1},\mu _{2}\right) }{3244+A_{2}^{\prime }\left(
p_{1},p_{2},\mu _{1},\mu _{2}\right) }  \nonumber \\
q_{2}^{\ast } &=&\frac{6561k\cos ^{2}\theta +B_{1}^{\prime }\left(
p_{1},p_{2},\mu _{1},\mu _{2}\right) }{52488+B_{2}^{\prime }\left(
p_{1},p_{2},\mu _{1},\mu _{2}\right) }  \label{21}
\end{eqnarray}%
where the damping parameters $A_{1}^{\prime },A_{2}^{\prime },B_{1}^{\prime
} $ and $B_{2}^{\prime }$ are given in appendix A. For unentangled initial
state, firms' payoffs at the subgame perfect Nash equilibrium point become%
\begin{eqnarray}
P_{A} &=&-\frac{(3-2p_{2})^{2}\{1+2p_{2}(-1+\mu _{2})\}^{2}}{%
24\{-3-6p_{2}(-1+\mu _{2})+4p_{2}^{2}(-1+\mu _{2})\}}  \nonumber \\
P_{B} &=&-\frac{(3-2p_{2})^{2}\{1+2p_{2}(-1+\mu
_{2})\}^{2}\{-3+2p_{2}(-3+2p_{2})(-1+\mu _{2})\}}{48[9+p_{2}(-3+2p_{2})%
\{10+2p_{2}(-3+2p_{2})(-1+\mu _{2})^{2}-9\mu _{2}\}]}  \nonumber \\
&&  \label{22}
\end{eqnarray}%
where we have considered decoherence only in the second evolution of the
game and\ have set $k=1$. It can easily be checked that for uncorrelated
channel, firms' moves for $p_{2}>0.5$ become negative and hence no Nash
equilibrium exits for $p_{2}>0.5$. On the other hand, it can be easily shown
that the Nash equilibrium exits for the entire range of values of
decoherence parameter, when the channel is correlated.
\begin{figure}[h]
\begin{center}
\begin{tabular}{ccc}
\vspace{-0.5cm}
\includegraphics[scale=0.4]{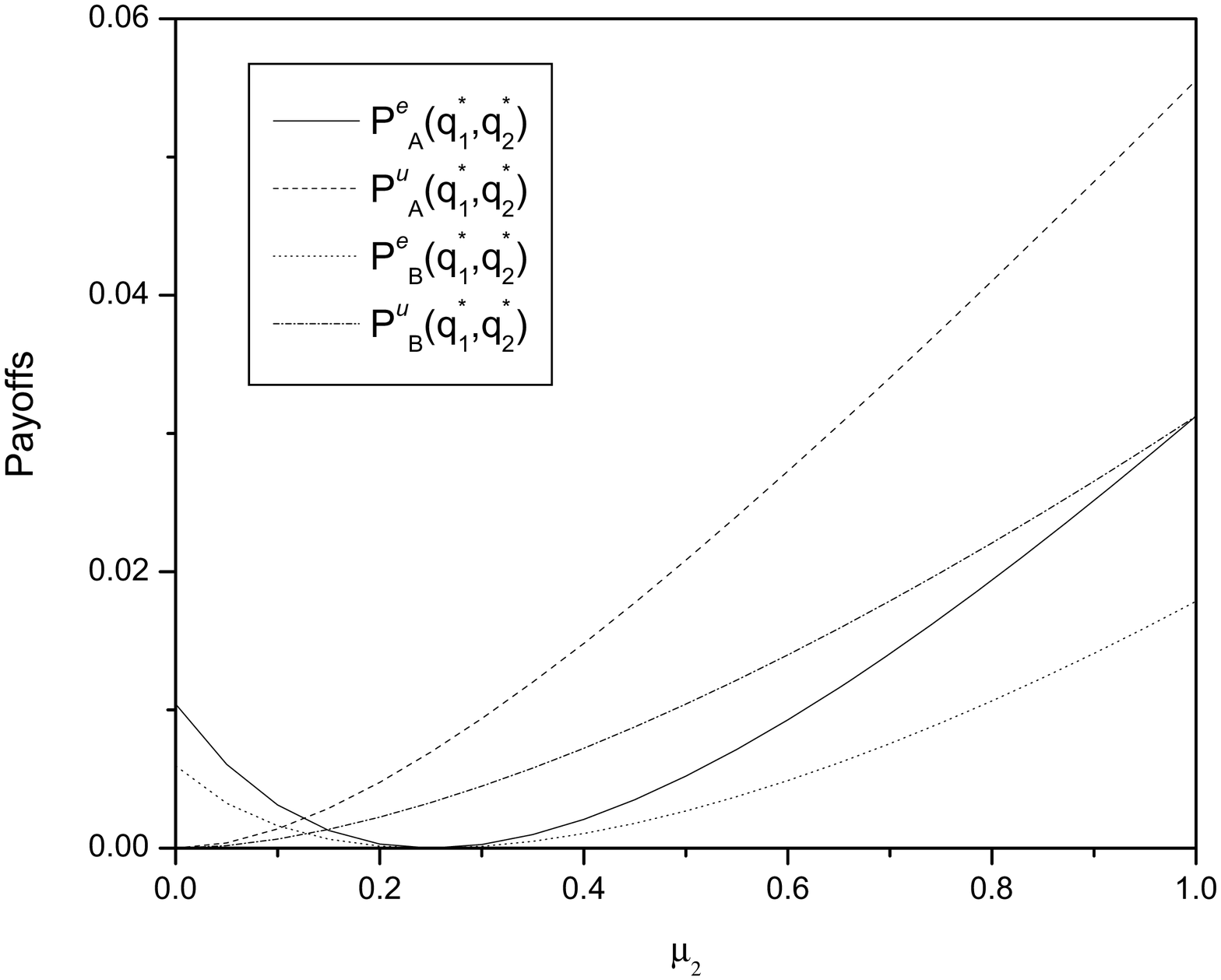}\put(-350,220) &  &
\end{tabular}%
\end{center}
\caption{The payoffs ($P_{A}(q_{1}^{\ast },q_{2}^{\ast })$ and $%
P_{B}(q_{1}^{\ast },q_{2}^{\ast })$) are plotted at the subgame perfect Nash
equilibrium point against the memory parameter $\protect\mu _{2}$ both for
entangled and unentangled initial states under the action of depolarizing
channel, with the other parameters$\ p_{1}=\protect\mu _{1}=0$, $k=1$ and $\
p_{2}=0.5$}
\end{figure}
The effect of quantum memory on firms' payoffs for unentangled initial state
under depolarizing channel is shown in figure $3$. The dashed line
represents firm $A$'s payoff and dashed-dotted line represnets firm $B$'s
payoff for unentangled initial state. The payoffs grow up with increasing
value of memory parameter and the game remains the first mover advantage
game.

When the initial state is maximally entangled and only the second decohering
process is taken into account, the moves and payoffs of the firms for $k=1$
become%
\begin{eqnarray}
q_{1}^{\ast } &=&\frac{1}{2}+\frac{3}{4\{-3+2p_{2}(-3+2p_{2})(-1+\mu _{2})\}}
\nonumber \\
q_{2}^{\ast } &=&\frac{\{-3+2p_{2}(-3+2p_{2})(-1+\mu
_{2})\}\{-3+4p_{2}(-3+2p_{2})(-1+\mu _{2})\}}{63+8p_{2}(-3+2p_{2})%
\{-9+2p_{2}(-3+2p_{2})(-1+\mu _{2})\}(-1+\mu _{2})}  \nonumber \\
&&  \label{23}
\end{eqnarray}%
\begin{eqnarray}
P_{A} &=&-\frac{\{3-4p_{2}(-3+2p_{2})(-1+\mu _{2})\}^{2}}{%
96\{-3+2p_{2}(-3+2p_{2})(-1+\mu _{2})\}}  \nonumber \\
P_{B} &=&-\frac{\{3-4p_{2}(-3+2p_{2})(-1+\mu
_{2})\}^{2}\{-3+2p_{2}(-3+2p_{2})(-1+\mu _{2})\}}{24[63+8p_{2}(-3+2p_{2})%
\{-9+2p_{2}(-3+2p_{2})(-1+\mu _{2})\}(-1+\mu _{2})]}  \nonumber \\
&&  \label{24}
\end{eqnarray}%
From equation (\ref{23}) one can easily check that no Nash equilibrium
exists for decoherence parameter $p_{2}>0.32$ when the channel is
uncorrelated. However, the presence of quantum correlations ensure the
presence of Nash equilibrium for the entire range of decoherence parameter.
It can be checked that for a given value of memory parameter the payoffs
decrease when plotted as function of decoherence parameter and the game is
always first mover advantage game. The dependence of payoffs for the
maximally entangled initial state on the memory parameter is shown in figure
$3$. The solid line in the figure represents firm $A$'s payoff and the
dotted line represents firm $B$'s payoff for the maximally entangled initial
state. It can be seen that the game is no-payoff game around $\mu _{2}=0.25$%
. For other values of memory parameter, the game is first mover advantage
game and the firms are better off when the channel is fully correlated.
\begin{figure}[h]
\begin{center}
\begin{tabular}{ccc}
\vspace{-2cm} \includegraphics[scale=0.4]{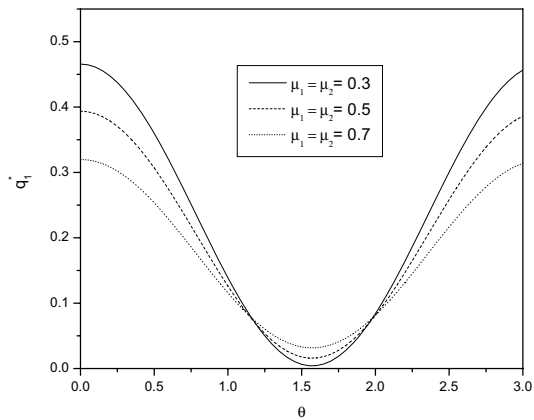}\put(-350,220) &
&
\end{tabular}%
\end{center}
\caption{The payoffs ($P_{A}(q_{1}^{\ast },q_{2}^{\ast })$ and $%
P_{B}(q_{1}^{\ast },q_{2}^{\ast })$) are plotted at the subgame perfect Nash
equilibrium point against the memory parameter $\protect\mu _{2}$ both for
entangled and unentangled initial states under the action of depolarizing
channel, with the other parameters$\ p_{1}=\protect\mu _{1}=0$, $k=1$ and $\
p_{2}=0.5$}
\end{figure}

To analyze the effect of entanglement in the initial state of the game on
the Nash equilibrium, we plot the firms' moves $q_{1}^{\ast }$, $q_{2}^{\ast
}$ in figures $4$ and $5$, respectively. From figure 4, one can see that the
move of the leader firm is positive for the whole range of entanglement
angle. However, for memory parameters $\mu _{1}=\mu _{2}<0.5$, the move $%
q_{2}$ of the follower firm is negative for a particular range of values of
entanglement parameter $\theta $ as can be seen from figure $5$.
\begin{figure}[h]
\begin{center}
\begin{tabular}{ccc}
\vspace{-2cm} \includegraphics[scale=0.4]{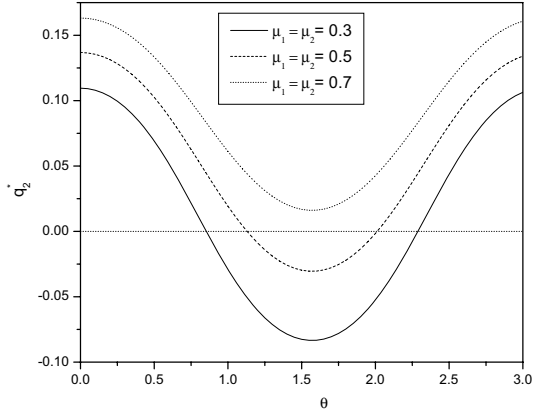}\put(-350,220) &
&
\end{tabular}%
\end{center}
\caption{The move of firm $B$ at the subgame perfect Nash equilibrium point
under the influence of depolarizing channel\ is plotted against the
entanglement angle $\protect\theta $ with other parameters$\
p_{1}=p_{2}=0.25 $, $k=1$.}
\end{figure}
The negative value of $q_{2}$ shows that no Nash equilibrium of the game
exists in this range of values of $\theta $. On the other hand, the
existence of Nash equilibrium is ensured for the whole range of values of
the entanglement parameter for highly correlated channel.
\begin{figure}[h]
\begin{center}
\begin{tabular}{ccc}
\vspace{-2cm} \includegraphics[scale=0.4]{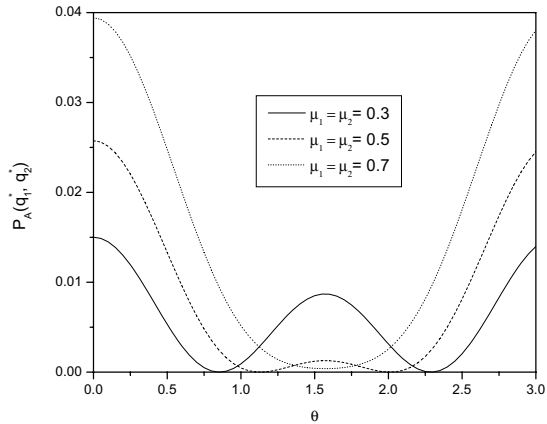}\put(-350,220) &
&
\end{tabular}%
\end{center}
\caption{The payoff of firm $A$ at the subgame perfect Nash equilibrium
point under the influence of depolarizing channel\ is plotted against the
entanglement angle $\protect\theta $ with other parameters$\
p_{1}=p_{2}=0.25 $, $k=1$.}
\end{figure}
The payoffs of firms as function of entanglement parameter are plotted in
figures $6$ and $7$. A local maximum exists at $\theta =\pi $ for smaller
values of memory parameters which disappears for the values of memory
parameters $>0.5$.
\begin{figure}[h]
\begin{center}
\begin{tabular}{ccc}
\vspace{-2cm} \includegraphics[scale=0.4]{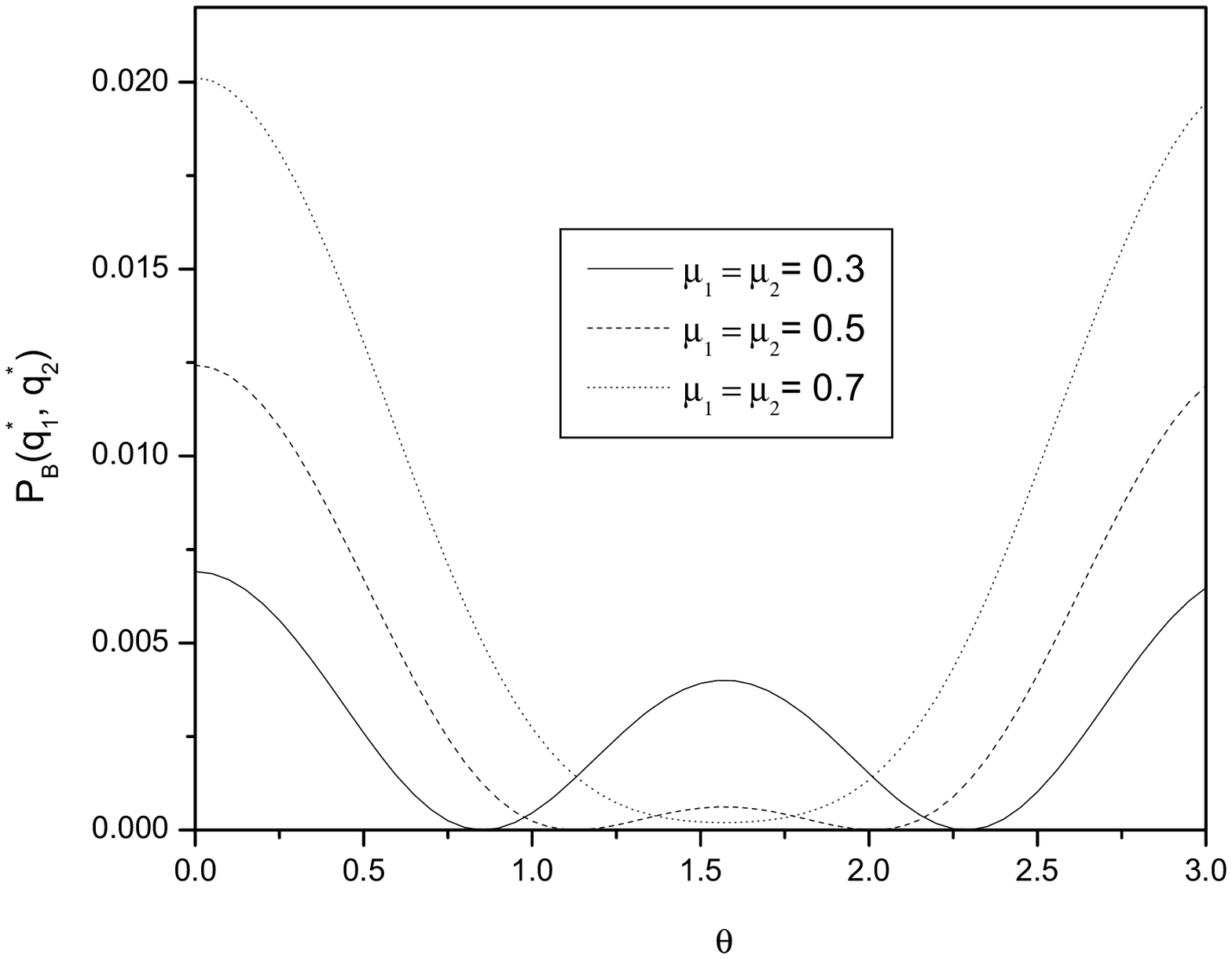}\put(-350,220) &
&
\end{tabular}%
\end{center}
\caption{The payoff of firm $B$ at the subgame perfect Nash equilibrium
point under the influence of depolarizing channel\ is plotted against the
entanglement angle $\protect\theta $ with the following values of the other
parameters$\ p_{1}=p_{2}=0.25$, $k=1$.}
\end{figure}

\subsection{Correlated phase damping channel}

The diagonal elements of the final density matrix of the game when it
evolves under the action of correlated phase damping channel are given by%
\begin{eqnarray}
\rho _{11}^{\prime } &=&q_{12}(\cos ^{2}\theta +q_{1}q_{2}\sin ^{2}\theta )
\nonumber \\
\rho _{22}^{\prime } &=&q_{12}(q_{2}\cos ^{2}\theta +q_{1}\sin ^{2}\theta )
\nonumber \\
\rho _{33}^{\prime } &=&q_{12}(q_{1}\cos ^{2}\theta +q_{2}\sin ^{2}\theta )
\nonumber \\
\rho _{44}^{\prime } &=&q_{12}(q_{1}q_{2}\cos ^{2}\theta +\sin ^{2}\theta )
\label{25}
\end{eqnarray}%
The payoffs of firms in quantum Stackelberg duopoly game depend only on the
diagonal elements of the final density matrix as can be seen from equations (%
\ref{11} and \ref{12}). It is clear from equation (\ref{25}) that the
diagonal elements of the final density matrix are independent of the
decoherence parameters as well as from the memory parameters. Therefore, the
correlated phase damping channel does not effect the outcome of the game.

\section{Conclusions}

We study the influence of entanglement and correlated noise on the quantum
Stackelberg duopoly game by considering the time correlated amplitude
damping, depolarizing and phase damping channels using Kraus operator
formalism. We have shown that in different entangling regions the follower
advantage can be enhanced or weakened due to the existence of the initial
state entanglement influenced by different correlated noise channels. The
problem of nonexistence of the subgame perfect Nash equilibrium in various
regions due to the presence of decoherence is resolved by quantum memory.

It is shown that under the action of amplitude damping channel, for
initially unentangled state, the presence of quantum memory results into a
critical point at the subgame perfect Nash equilibrium. The firms are
equally benefited at this point and the leader advantage vanishes. Beyond
this critical point, the Nash equilibrium of the game gives higher payoff to
the follower firm as a result of quantum memory, that is, the game converts
from leader advantage to the follower advantage game. Quantum memory, in the
case of amplitude damping channel, favors the follower firm in a particular
range of values of entanglement parameter at the subgame perfect Nash
equilibrium. In this range of values of entanglement parameter, the leader
firm get worse off and the follower firm is better off (see figure 2). It is
also observed that quantum memory validates the existence of Nash
equilibrium for the whole range of entanglement angle and decoherence
parameter. It is also shown that quantum memory in the case of phase damping
channel has no effect on the subgame perfect Nash equilibrium and thus does
not change the outcome of the game.

In the case of depolarizing channel quantum memory and entanglement in the
initial state influences the firms' payoffs at the subgame perfect Nash
equilibrium strongly in a way different from amplitude damping channel. It
is seen that for $p>0.32$, no Nash equilibrium exists in case of unentangled
initial state. Whereas the presence of entanglement in the initial state
extends the span of decoherence parameter $p$ from $0-0.32$ to $0-0.5$ for
the existence of subgame perfect Nash equilibrium in the absence of quantum
memory. On the other hand, we have observed that in the presence of memory,
the subgame perfect Nash equilibrium exists for the entire range of
decoherence parameter in both of the situations (for entangled and
unentangled initial states). Similarly, it has been shown that memory has a
striking effect that there exists a Nash equilibrium of the game for the
entire range of entanglement parameter as well. Whereas, in the absence of
memory, the noisy environment limits the subgame perfect Nash equilibrium to
exist in a particular range of entanglement angle. In addition, a local
maximum in payoffs is observed for small values of quantum memory parameters
$\mu _{1}$, $\mu _{2}$. For highly correlated channel this local maximum
disappears and the payoffs reduce to zero. Unlike amplitude damping channel,
the correlated depolarizing channel does not give rise to a critical point
at the subgame perfect Nash equilibrium and as a result, the game always
remains a leader advantage game.

\section{Acknowledgment}

We would like to thank the referees of Journal of Physics A:\ Mathematical
and Theoretical, for their substantial contribution to the improvement of
the manuscript by their useful suggestions. One of the authors (Salman Khan)
is thankful to World Federation of Scientists for partially supporting this
work under the National Scholarship Program for Pakistan\textbf{.}

\begin{center}
{\LARGE Appendix A}
\end{center}

The damping functions $A_{i}$ in equation (\ref{15}) are given as%
\begin{eqnarray}
A_{1} &=&\frac{1}{2}[-k-4(-1+p_{1})(-1+p_{2})p_{2}(-1+\mu _{2})  \nonumber \\
&&-2kp_{2}(p_{2}+\mu _{2}-p_{2}\mu _{2})-2kp_{1}[\mu
_{1}+p_{2}\{2-p_{2}(-2+\mu _{1})  \nonumber \\
&&\times (-1+\mu _{2})-4\mu _{2}+\mu _{1}(-2+3\mu _{2})\}]  \nonumber \\
&&-k\cos 2\theta +4(-1+p_{1})(-1+p_{2})p_{2}(-1+\mu _{2})\cos 2\theta
\nonumber \\
&&+2kp_{2}(p_{2}+\mu _{2}-p_{2}\mu _{2})\cos 2\theta +2kp_{1}[\mu
_{1}+p_{2}\{2-p_{2}(-2+\mu _{1})  \nonumber \\
&&\times (-1+\mu _{2})-4\mu _{2}+\mu _{1}(-2+3\mu _{2})\}]\cos 2\theta
]-2p_{1}(-1+\mu _{1})  \nonumber \\
&&\times \lbrack 2+p_{1}[-2+k\{-1+p_{2}(2-3\mu _{2})+p_{2}^{2}(-1+\mu _{2})\}
\nonumber \\
&&-4p_{2}(-1+\mu _{2})+2p_{2}^{2}(-1+\mu _{2})]+4p_{2}(-1+\mu _{2})
\nonumber \\
&&-2p_{2}^{2}(-1+\mu _{2})]\sin ^{2}\theta
\end{eqnarray}%
\begin{eqnarray}
A_{2} &=&-4(1+k)p_{2}(-1+\mu _{2})-4(-1+p_{1})p_{1}(-1+\mu _{1})  \nonumber
\\
&&\times \lbrack -2+k\{-1+p_{2}(2-3\mu _{2})+p_{2}^{2}(-1+\mu _{2})\}
\nonumber \\
&&-4p_{2}(-1+\mu _{2})+2p_{2}^{2}(-1+\mu _{2})]\sin ^{2}\theta
\end{eqnarray}%
The damping functions $B_{i}$ in equation (\ref{15}) are given as%
\begin{eqnarray}
B_{1} &=&-2+\frac{1}{8}k[1+2p_{2}(p_{2}+\mu _{2}-p_{2}\mu _{2})  \nonumber \\
&&+2p_{1}[\mu _{1}+p_{2}\{2-4\mu _{2}+\mu _{1}(-2+3\mu _{2})\}]  \nonumber \\
&&+[1-2[p_{2}(p_{2}+\mu _{2}-p_{2}\mu _{2})+p_{1}\{\mu _{1}+p_{2}(2-2\mu _{1}
\nonumber \\
&&-4\mu _{2}+3\mu _{1}\mu _{2})\}]]\cos 2\theta ]-\frac{1}{2}p_{1}(-1+\mu
_{1})[k(-4+3p_{1})  \nonumber \\
&&\times \{-1+p_{2}(2-3\mu _{2})+p_{2}^{2}(-1+\mu
_{2})\}+6(-1+p_{1})\{-1-2p_{2}  \nonumber \\
&&\times (-1+\mu _{2})+p_{2}^{2}(-1+\mu _{2})\}]\sin ^{2}\theta -\frac{1}{2}%
p_{2}(-1+\mu _{2})  \nonumber \\
&&\times \lbrack 4(1+k)+[2(-1+p_{2})+p_{1}\{2+p_{2}(-2+k(-2+\mu
_{1}))\}]\sin ^{2}\theta ]  \nonumber \\
&&
\end{eqnarray}%
\begin{eqnarray}
B_{2} &=&-16+2[2\{-1-(1+k)p_{2}(-1+\mu _{2})\}+p_{1}^{2}(-1+\mu _{1})
\nonumber \\
&&\times \lbrack (2+k)(-1+p_{2})^{2}+p_{2}\{4+3k-(2+k)p_{2}\}\mu _{2}]
\nonumber \\
&&+p_{1}(-1+\mu _{1})[-(2+k)(-1+p_{2})^{2}+p_{2}\{-4-3k  \nonumber \\
&&+(2+k)p_{2}\}\mu _{2}]+(-1+p_{1})p_{1}(-1+\mu _{1})[-(2+k)(-1+p_{2})^{2}
\nonumber \\
&&+p_{2}\{-4-3k+(2+k)p_{2}\}\mu _{2}]\cos 2\theta ][\{1+(1+k)p_{2}  \nonumber
\\
&&\times (-1+\mu _{2})\}\cos ^{2}\theta +[1+(1+k)p_{2}(-1+\mu
_{2})+p_{1}(-1+\mu _{1})  \nonumber \\
&&\times \lbrack (2+k)(-1+p_{2})^{2}+p_{2}\{4+3k-(2+k)p_{2}\}\mu _{2}]
\nonumber \\
&&+p_{1}^{2}(-1+\mu _{1})[-(2+k)(-1+p_{2})^{2}+p_{2}\{-4-3k  \nonumber \\
&&+(2+k)p_{2}\}\mu _{2}]\sin ^{2}\theta ]-[-2(-1+p_{1})[p_{1}(-1+\mu _{1})
\nonumber \\
&&\times \{-1+(-2+p_{2})p_{2}(-1+\mu _{2})\}+(-1+p_{2})p_{2}(-1+\mu _{2})]
\nonumber \\
&&-k[1+p_{2}(p_{2}+\mu _{2}-p_{2}\mu _{2})+p_{1}^{2}(-1+\mu
_{1})\{-(-1+p_{2})^{2}  \nonumber \\
&&+(-3+p_{2})p_{2}\mu _{2}\}+p_{1}[\mu _{1}+p_{2}\{2-p_{2}(-2+\mu
_{1})(-1+\mu _{2})  \nonumber \\
&&-4\mu _{2}+\mu _{1}(-2+3\mu _{2})\}]]+(-1+p_{1})[2\{p_{1}(-1+\mu _{1})
\nonumber \\
&&\times \{-1+(-2+p_{2})p_{2}(-1+\mu _{2})\}+(-1+p_{2})p_{2}(-1+\mu _{2})\}
\nonumber \\
&&+k[(-1+p_{2})\{-1+p_{2}(-1+\mu _{2})\}+p_{1}(-1+\mu _{1})\{-(-1+p_{2})^{2}
\nonumber \\
&&+(-3+p_{2})p_{2}\mu _{2}\}]]\cos 2\theta ][p_{2}\{-2-(2+k)p_{2}(-1+\mu
_{2})  \nonumber \\
&&+(2+k)\mu _{2}\}\cos ^{2}\theta +[2p_{1}[-1+\mu _{1}+p_{1}(-1+\mu _{1})
\nonumber \\
&&\times \{-1+(-2+p_{2})p_{2}(-1+\mu _{2})\}-p_{2}\{1+(-2+p_{2})\mu
_{1}\}(-1+\mu _{2})]  \nonumber \\
&&+k[1+p_{1}[-2+\mu _{1}+p_{1}(-1+\mu
_{1})\{-(-1+p_{2})^{2}+(-3+p_{2})p_{2}\mu _{2}\}  \nonumber \\
&&+p_{2}\{2-2\mu _{2}+\mu _{1}(-2+p_{2}+3\mu _{2}-p_{2}\mu _{2})\}]]]\sin
^{2}\theta ]
\end{eqnarray}%
The damping functions $A_{i}^{\prime }$ in equation (\ref{21}) are given as

\begin{eqnarray}
A_{1}^{\prime } &=&4(2+k)p_{1}(-3+2p_{1})(-1+\mu _{1})\{-9+8p_{2}(-3+2p_{2})
\nonumber \\
&&\times (-1+\mu _{2})\}-36(2+k)p_{2}(-3+2p_{2})(-1+\mu _{2})  \nonumber \\
&&+\frac{9}{2}k[9+\{9-24p_{2}+8p_{1}(-3+4p_{2})\}\cos 2\theta ]
\end{eqnarray}%
\begin{eqnarray}
A_{2}^{\prime } &=&8(2+k)[p_{1}(-3+2p_{1})(-1+\mu _{1})\{-9+8p_{2}(-3+2p_{2})
\nonumber \\
&&\times (-1+\mu _{2})\}-9p_{2}(-3+2p_{2})(-1+\mu _{2})]
\end{eqnarray}%
The damping functions $B_{i}^{\prime }$ in equation (\ref{21}) are given as%
\begin{eqnarray}
B_{1}^{\prime }
&=&8(2+k)^{2}(3-2p_{2})^{2}p_{2}^{2}\{9-8p_{1}(-3+2p_{1})(-1+\mu _{1})\}^{2}
\nonumber \\
&&\times \lbrack k\{-9+4p_{1}(-3+2p_{1})(-1+\mu _{1})\}+8p_{1}(-3+2p_{1})
\nonumber \\
&&\times (-1+\mu _{1})-k(-3+4p_{1})(-3+4p_{2})\cos 2\theta ]+18(2+k)
\nonumber \\
&&\times p_{2}(-3+2p_{2})\{-9+8p_{1}(-3+2p_{1})(-1+\mu _{1})\}(-1+\mu _{2})
\nonumber \\
&&\times \lbrack (2+k)\{9-8p_{1}(-3+2p_{1})(-1+\mu _{1})\}+k(-3+4p_{1})
\nonumber \\
&&\times (-3+4p_{2})\cos 2\theta ]
\end{eqnarray}%
\begin{eqnarray}
B_{2}^{\prime } &=&-72(2+k)p_{2}(-3+2p_{2})[4\{-9+2p_{1}(-3+2p_{1})(-1+\mu
_{1})\}  \nonumber \\
&&+k\{9+4p_{1}(-3+2p_{1})(-1+\mu _{1})\}]\{-9+8p_{1}  \nonumber \\
&&\times (-3+2p_{1})(-1+\mu _{1})\}(-1+\mu _{2})+16(2+k)^{2}(3-2p_{2})^{2}
\nonumber \\
&&\times p_{2}^{2}\{9-8p_{1}(-3+2p_{1})(-1+\mu _{1})\}^{2}(-1+\mu _{2})^{2}
\nonumber \\
&&+81[k^{2}\{-81+8p_{1}(-3+2p_{1})\{9+2p_{1}(-3+2p_{1})(-1+\mu _{1})\}
\nonumber \\
&&\times (-1+\mu _{1})\}+64p_{1}(-3+2p_{1})\{-9+p_{1}(-3+2p_{1})  \nonumber
\\
&&\times (-1+\mu _{1})\}(-1+\mu _{1})+16kp_{1}(-3+2p_{1})\{-9+4p_{1}
\nonumber \\
&&\times (-3+2p_{1})(-1+\mu _{1})\}(-1+\mu _{1})+k^{2}(3-4p_{1})^{2}
\nonumber \\
&&\times (3-4p_{2})^{2}\cos ^{2}2\theta ]
\end{eqnarray}

\end{document}